\documentclass[aps,prl,twocolumn,showpacs,superscriptaddress,groupedaddress]{revtex4}  % for review and submission
\usepackage{graphicx}  % needed for figures
\usepackage{dcolumn}   % needed for some tables
\usepackage{bm}        % for math
\usepackage{amssymb}   % for math
\usepackage{physics}
\usepackage{relsize}

\usepackage{multibib}
\usepackage{hyperref} 

\usepackage[utf8]{inputenc}
\usepackage{amsmath,tikz,subcaption}

\begin{document}
\title{Phase locking and noise driven dynamics in a Josephson junction electronic analog}
\author{Aeron McConnell$^{*}$, Sara Idris$^{*}$,  Brian Opatosky and Fran\c{c}ois Amet$^{\dagger}$}

\affiliation{Department of Physics and Astronomy, Appalachian State University}

\date{\today}

\begin{abstract}
We present an electronic circuit whose dynamical properties emulate those of a resistively and capacitively shunted Josephson junction (RCSJ). We show how it reproduces the switching properties of a shunted junction and its dependence on the quality factor. A thermal noise source is then used to characterize the temperature dependence of the phase dynamics. In the presence of an AC drive, phase locking is observed at integer and rational multiples of the drive frequency, and it competes with chaotic behavior when the quality factor of the junction exceeds unity. We characterize the stability of phase-locked and chaotic states in the presence of thermal noise. 

\end{abstract}

\pacs{}
\maketitle
\def\thefootnote{*}\footnotetext{These authors contributed equally to this work}
\def\thefootnote{$\dagger$}\footnotetext{Email: ametf@appstate.edu}

\section{Introduction}

Josephson junctions consist of two superconducting electrodes separated by a weak link \cite{JosephsonRMP1964, Tinkham}. The supercurrent flowing through a junction depends on the superconducting phase difference $\phi$ between the two electrodes. The electronic properties of a junction largely result from the dynamical properties of $\phi(t)$, which are equivalent to those of a driven nonlinear oscillator. In fact, within the resistively and capacitively shunted junction model (RCSJ) \cite{mccumber_effect_1968,stewart_currentvoltage_1968}, $\phi$ follows a differential equation which is identical to that of a driven, damped, rigid pendulum. 

Electronic circuits relying on mainstream components can emulate the exact same dynamical system, and thus provide a convenient way to simulate the properties of Josephson junctions as predicted by the RCSJ model. Such Josephson junction analogues were proposed in Ref. \cite{hamilton_analog_1972,dhumieres_chaotic_1982,gundlach_analogue_2008,magerlein_accurate_1978,yagi_precision_2008,henry_simple_1981, russer_influence_2003,blackburn_circuit_2007} and typically rely on voltage-controlled oscillators. Prior work also demonstrated how analog circuits can help characterize noise-driven dynamics \cite{luchinsky_irreversibility_1997}, such as a thermally activated escape rate from a potential well \cite{teitsworth_scaling_2019}. However, to our knowledge this approach has not been used to investigate the effect of thermal noise on the phase dynamics of a Josephson junction.

In this work, we probe some of the common properties of Josephson junctions using an analog circuit. We first determine the switching properties of the analog junction and its dependence on the quality factor.  The addition of thermal noise to the circuit allows the observation of phase diffusion in overdamped junctions, as well as the thermally activated premature switching of underdamped junctions.  In the presence of an AC drive, we observe the AC Josephson effect \cite{Shapiro1963, Tinkham} and its dependence on the Q factor of the junction, as well as the power and frequency of the AC drive. Finally, we discuss the onset of chaos in such systems and the thermal stability of phase-locked and chaotic states. 

%Recent years have seen a renewed interest in the phase dynamics of more complex Josephson junctions. For example, the AC Josephson effect provides a way to determine the current phase relation of Josephson junctions relying on topologically non-trivial materials. More recently, the DC and AC Josephson effects were used to probe the properties of multiterminal Josephson junctions, 

\section{Description of the circuit}
\subsection{Main circuit}

\begin{figure}
    \centering
    \includegraphics[width=3.375 in,keepaspectratio]{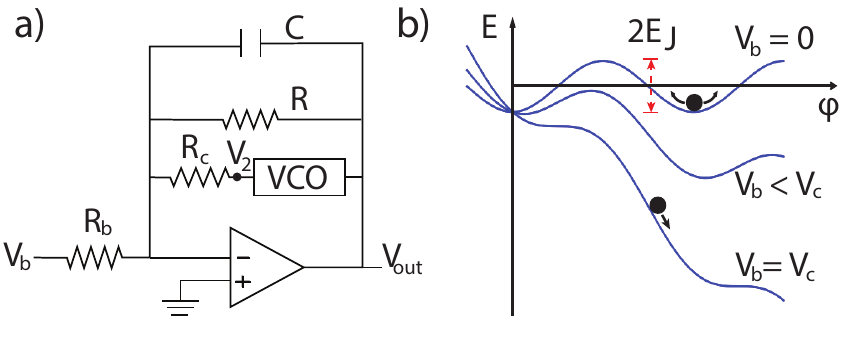}
    \caption{A simplified representation of the circuit is shown in (a) with the most relevant components labeled. (b) gives an idealized diagram of washboard potentials at varying bias currents. }
    \label{fig:figure1}
\end{figure}

The circuit, inspired by Ref.\cite{blackburn_circuit_2007}, relies on a voltage controlled oscillator (VCO) whose sinusoidal output $V_{2}$ has an amplitude $\alpha$, and a frequency which is proportional to the input voltage ($V_{out}$ on Figure 1): $V_{2}=\alpha \sin(2\pi k\int V_{ out}dt)$. A constant input voltage $V_{out}$ therefore relies on a sinusoidal output of frequency $k V_{out}$.

We characterized the VCO in greater detail in the supplementary information. If we define $\dot{\phi}\equiv 2{\pi}k V_{out}$, we obtain the following differential equation for $\phi$:

\begin{equation}\label{equation 1}
\ddot {\phi} + \frac{\omega_{0}}{Q} \dot {\phi} + \omega_{0}^{2}\sin{\phi}=\omega_{0}^{2}\frac{V_{b}}{V_{c}}
\end{equation}

Here the junction's frequency is defined as 
${\omega}_o^2 \equiv \frac{2{\pi}{\alpha}k}{R_cC}$, the quality factor $Q\equiv RC\omega_{0}$, and the critical voltage $V_{c}\equiv\frac{-\alpha R_b}{R_c}$. A more detailed derivation of this result is provided in the supplementary information and we show how to determine $Q$ and $\omega_{0}$ experimentally. 

\begin{figure*}
    \centering
    \includegraphics[width=\textwidth]{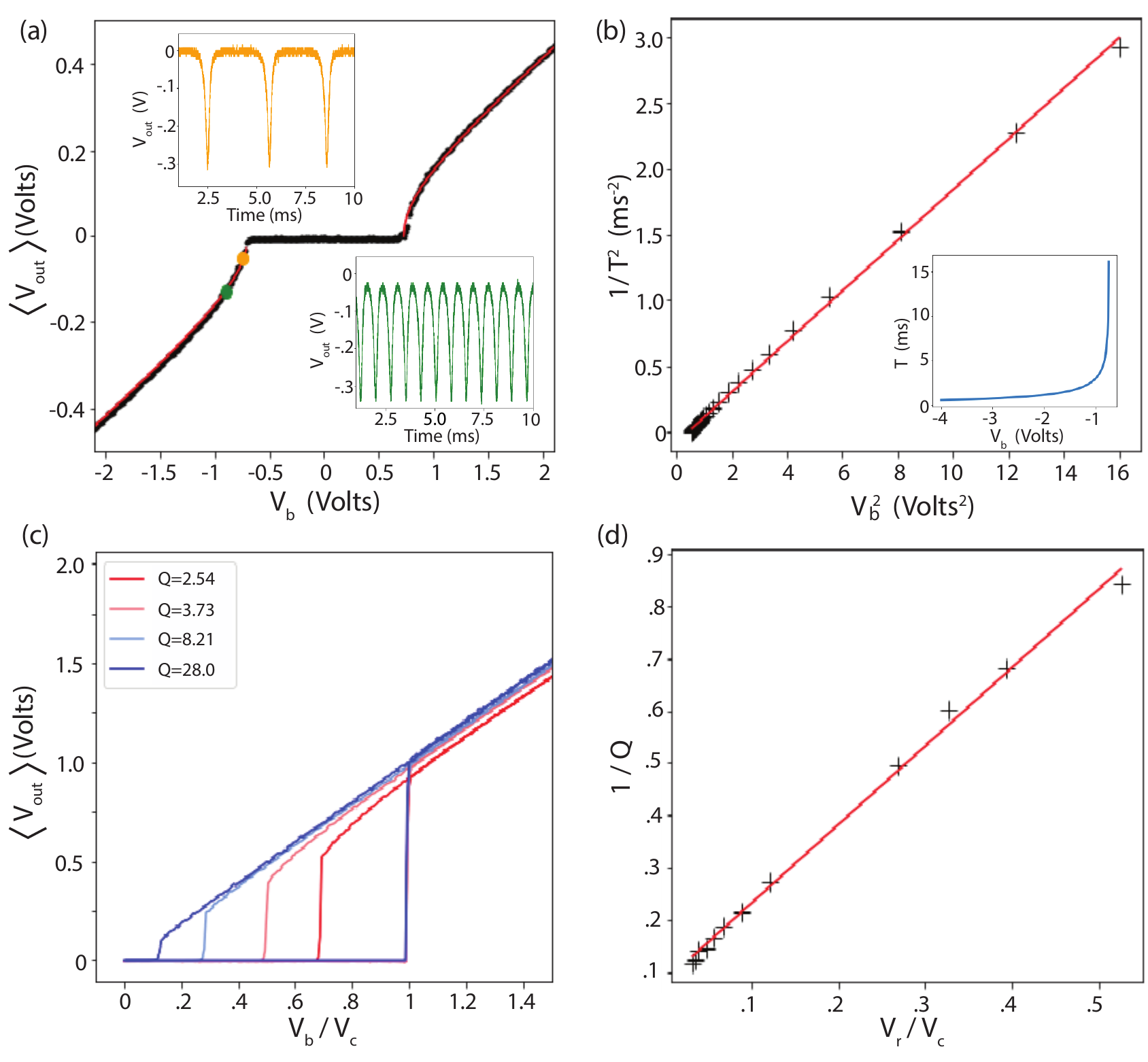}
    \caption{Switching Dynamics. (a) $\langle V_{out}\rangle$ measured as a function of the DC bias voltage for an analog junction with Q=0.6. The red curve is a fit to the RCSJ model prediction (Eq. 3). Insets: $V_{out}(t)$ before time averaging measured at bias values shown by yellow and green dots on the main curve. The period of oscillation clearly increases closer to the critical voltage. 
    (b) $1/T^2$ plotted as a function of $V_b^2$, which is expected to be linear according to equation 4. A fit, shown in red and relying on equation 4, yields a junction period $T_{0}=2.1$ms. Inset: $T(V_{b})$ using the same data. (c) I-V curves measured for different quality factors. The retrapping voltage decreases at larger Q values. (d) Given the RCSJ model's prediction that $V_r/V_c\propto 1/Q$ the values of $1/Q$ versus $V_r/V_c$ were plotted for 12 different $Q$ values to show this proportionality.}
    \label{fig:figure2}
\end{figure*}

\begin{figure*}
    \centering
    \includegraphics[width=\textwidth]{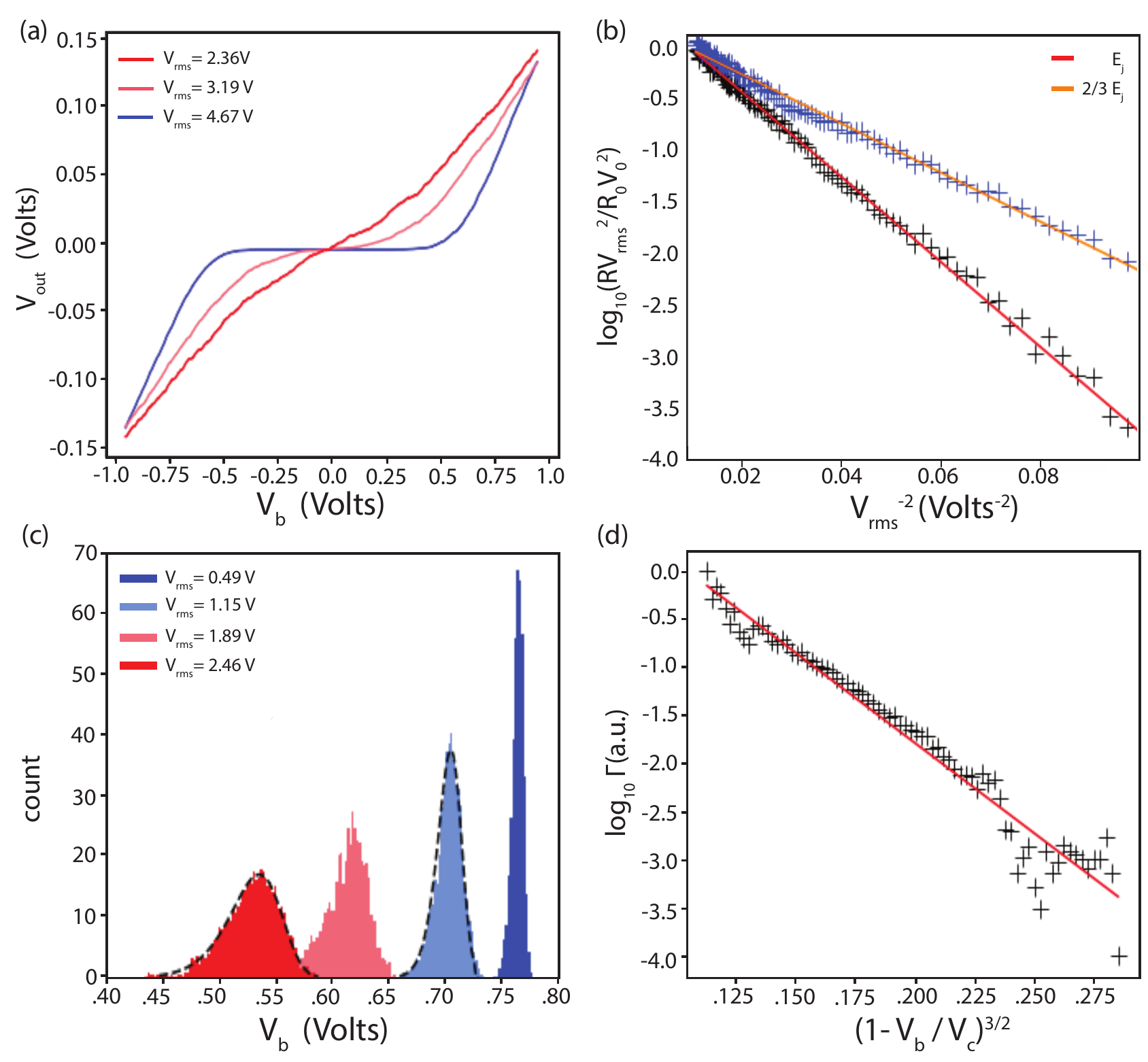}
    \caption{Thermal Noise. (a) Increasing the RMS voltage of the noise source causes the superconducting region to start tilting and causes the junction to become slightly resistive even below the critical voltage. At high enough RMS voltages the superconducting behavior is entirely washed out by the thermal noise. (b) Plotting $\ln(\tilde{R}V_{rms}^2)$ versus $V_{rms}^{-2}$ we expect to see linear relationships with slopes proportional to $E_j$. (c) As the thermal noise RMS voltage is increased the probability distribution of the switching voltage becomes more broad and the mean drops towards lower bias. (d) By considering $ \Delta U=2E_{J}(1- V_{b}/V_{c})^{\frac{3}{2}}$, where the switching rate is proportional to $e^{-\Delta U/kT}$, we see that plotting $\log(\Gamma)$ versus $(1-V_b/V_c)^{3/2}$ gives a linear relationship which is confirmed in this plot.}
    \label{fig:figure3}
\end{figure*}
This equation is similar to the familiar RCSJ model for the phase of a current-biased Josephson junction \cite{stewart_currentvoltage_1968, mccumber_effect_1968, Tinkham}. The effective potential associated with the dynamical properties of this phase is:

\begin{align}
U(\phi)=-E_{J}\left(\cos(\phi)+\frac{V_{b}}{V_c}\phi\right)
\end{align}

Here, $E_{J}$ is the analogue of the Josephson energy in the RCSJ model and is defined as $E_J\equiv \frac{\alpha}{2{\pi}k\,R_c}$. For our circuit, this quantity is on the order of 80 nJ, which is of course many orders of magnitude greater than the energy scale for a real junction. Details on the derivation of this result are provided in the supplementary information. 

Note that the bias voltage $V_{b}$ plays the role of what is usually the current bias in the RCSJ model. In what follows, we will therefore refer to switching, retrapping and critical voltages instead of currents. Curves representing $V_{out}(V_{b})$ will be referred to as I-V curves to follow the standard terminology. 

In our setup, the bias voltage is the sum of three components: DC and AC voltages, as well as a random voltage source that emulates thermal noise and is described later in the paper. 
\subsection{Switching dynamics}

\par

Figure 2a shows an example of an I-V curve obtained when the analog junction is nearly overdamped with a Q factor of 0.6. Here, $V_{out}$ is time-averaged by a low pass filter of time constant 82 ms. As the bias exceeds the critical voltage, $V_{out}$ becomes finite, which corresponds to the normal state in the RCSJ model. In the overdamped case, the junction's I-V curve is not hysteretic, so the switching and retrapping voltages are identical \cite{Tinkham}. The red curve corresponds to a fit to the RCSJ model's prediction in the overdamped case:
\begin{equation}
    V_{out}=\pm\frac{f_{0}\,Q}{k}\sqrt{\left(\frac{V_{b}}{V_{c}}\right)^{2}-1}
\end{equation}

The two insets in Figure 2a represent $V_{out}\propto \dot{\phi}$ before time averaging. As the bias exceeds the critical voltage, the phase runs down the tilted washboard potential with an angular velocity that decreases whenever $\phi$ goes over a local maximum of $U(\phi)$. This results in an oscillation of $\dot{\phi}$ which is shown as an inset of Figure 2a. As the bias approaches the critical voltage, the period of those oscillations diverges, as shown in the insets 2a and 2b. Indeed, in the overdamped case, the period $T$ of those oscillations is expected to follow \cite{stewart_currentvoltage_1968, mccumber_effect_1968}:
\begin{equation}
    T=\frac{T_{0}}{Q\sqrt{\left(\mathlarger{\frac{V_{b}}{V_{c}}}\right)^{2}-1}}
\end{equation}

This trend is shown on Figure 2b where $T^{-2}$ is plotted as a function of $V_{b}^{2}$, showing a linear trend as expected from equation (9). The red fit corresponds to a period $T_{0}\approx 2.1$ ms (using Q$\approx$0.6). 

Changing the value of $R$ modifies the quality factor of the analog junction which can be tuned in the underdamped regime. As expected, the I-V curves become more hysteretic as the quality factor increases.
When $V_{b}>V_s\approx V_{c}$, the system starts to rapidly fall toward lower values of the washboard potential. However, as $V_{b}$ is decreased, a finite voltage is observed across the junction until the retrapping voltage $V_r<V_{s}$. Indeed, given the inertia of the system, the tilt of the washboard potential must be brought closer to the horizontal in order to stop the running phase. 
Figure 2c shows normalized I-V curves for four different quality factors. The retrapping current becomes increasingly small deeper in the underdamped regime.  Within the RCSJ model the ratio $V_{r}/V_{c}$ scales like $4/(Q\pi)$ for $Q>>1$ \cite{stewart_currentvoltage_1968}
; we therefore plot the ratio $V_{r}/V_{c}$ as a function of $1/Q$ in Figure 2d, and indeed observe a linear trend. 

\subsection{Effect of thermal noise}

\subsubsection{Effective temperature}

The effect of finite temperature on the phase dynamics can be emulated by applying gaussian white voltage noise to the input of the circuit. Using the standard expression for Johnson-Nyquist noise, this emulates an effective temperature: $k_{B}T_{eff}=\langle v_{rms}\rangle^2/4BR $

We numerically generate a random array of voltage values with a gaussian white noise distribution and, using a voltage output DAQ, add it to the DC bias voltage before feeding it to the input of the junction. The effective temperature is then simply varied by adjusting the RMS voltage of the white noise. Importantly, this method only works because the junction frequency $f_{0}$ is typically a few hundreds of Hz; much smaller than the repetition rate of the DAQ (10kHz). The spectral density of our noise source is therefore nearly flat up to a few kHz, which is sufficient in this low frequency setup to emulate thermal noise. This allows us to determine the temperature dependence of the phase dynamics in the overdamped and underdamped case. 

Around zero bias, phase slips between two local minima of the washboard potential can still occur at finite temperature with a probability that scales like the Boltzmann weight exp $(\mathlarger{-\Delta U/k_{B}T})$, with $\Delta U=2E_{J}$ \cite{kramers_brownian_1940}. This causes the phase to diffuse and a non-zero voltage to develop across the junction even when the bias voltage is below the critical voltage.
Figure 3a shows $V_{out}(V_{b})$ measured at three effective temperatures for an analog junction whose Q factor is 0.6.  The blue, pink, and red curves were obtained when the the RMS voltage of the noise source was 2.36 V, 3.19 V, and 4.67 V, respectively. A finite slope develops around zero bias and increases with temperature. At the highest effective temperature (red curve), the zero-voltage state effectively disappears. 

In the presence of phase diffusion, the zero bias resistance, which in our case corresponds to $dV_{out}/dV_{b}$ around $V_{b}=0$, is known to be thermally activated \cite{martinis_classical_1989} :

\begin{align}
     \tilde{R}\equiv \frac{dV_{out}}{dV_{b}} \Bigr\rvert_{V_{b}=0} \propto  \frac{1}{T}\exp\left(\frac{-2E_{j}}{k_{B}T}\right)
\end{align}

This implies that $\ln(\tilde{R}T)$ is proportional to $1/T$ with a slope proportional to $E_{j}$, which can be represented on an Arrhenius plot as shown on Figure 3b. $\langle V_{out}(V_{b})\rangle$ was measured in the presence of thermal voltage noise with varying RMS amplitude. The curves were numerically differentiated to get $dV_{out}/dV_{b}$ at zero bias. Since the effective temperature is proportional to $V_{rms}^2$, we plotted the decimal logarithm $\log(\tilde{R}V_{rms}^2)$ versus $V_{rms}^{-2}$, which shows an evident activated behavior in Figure 3b. To verify that the slope scaled according to $E_j$, we changed $E_j$ while keeping Q constant. When we change $E_j$ to 2/3 of its original value, we observe again a thermally activated behavior, but the slope is reduced by a comparable factor of $\approx$0.58. 

%The thermal noise bandwidth, B, can be determined experimentally using these values. From equation (5), $k_{B}T$ in $-2E_{j}/k_{B}T$ can be substituted for $\langle v_{rms}\rangle^2/4BR$. The expression becomes $-8E_{J}BR/V_{rms}^2$, where the slope is equal to $-8E_{J}BR$. Inserting the known values for the slope and 180\,$\Omega$ for R, the bandwidth in terms of $E_J$ is found to be $0.342/E_{J}$ for the circuit at $E_J$ and $0.296/E_{J}$ at two-thirds of $E_J$. When $0.578E_{J}$ is used to calculate the bandwidth of the modified circuit rather than $\frac{2}{3}E_{J}$, the result is again $0.342/E_{J}$. 

\begin{figure*}[t]
    \centering
    \includegraphics[width=\textwidth]{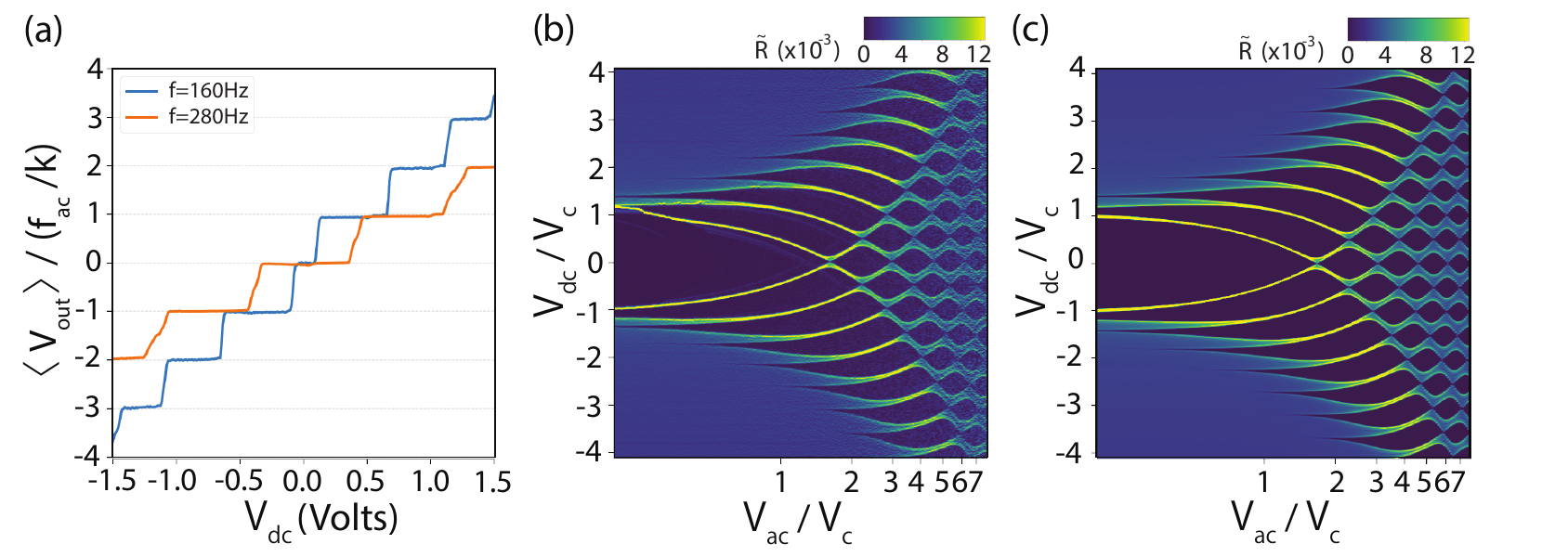}
    \caption{Inverse AC Josephson effect. (a) Plot of the average junction voltage, normalized in integral multiples of $f_{ac}/k$ versus $V_{dc}$. (b) Experimentally determined map of differential resistance $\widetilde{R}$ in $V_b$ versus $V_{ac}$ space; normalized in units of critical voltage $V_c$ ($\widetilde{R}$ is dimensionless). (c) Numerical simulation of part (b) via numerical integration of (\ref{equation 1}).
    }
    \label{fig:figure4}
\end{figure*}

In the underdamped case, a phase slip causes the phase to durably escape the zero-voltage state. The barrier height that determines the escape probability is bias voltage dependent and equal to $ \Delta U=2E_{J}(1- V_{b}/V_{c})^{\frac{3}{2}}$ \cite{buttiker_thermal_1983,silvestrini_current_1988,devoret_resonant_1987,teitsworth_scaling_2019}.  
To determine the switching voltage distribution, 4000 I-V curves were recorded for each effective temperature, wherein the bias voltage was ramped up to record the switching voltage. The distributions of switching voltage values were plotted for four effective temperatures in Figure 3c. As the thermal noise RMS amplitude is increased, the switching voltage distribution shifts towards lower bias voltages. Additionally, we observe a clear broadening of the switching voltage distribution with temperature, which concurs with the T$^{2/3}$ scaling predicted by the RCSJ model \cite{kurkijarvi_intrinsic_1972}.

The probability that switching occurs at a given voltage bias $V_{b}$ can be expressed in terms of the switching rate $\Gamma$, and the rate of change of the voltage bias $\dot{v}$ \cite{fulton_lifetime_1974, silvestrini_current_1988}:

\begin{align}
    P(V_{b}) =  \frac{\Gamma}{\dot{v}}\left(1- \int_{0}^{V_{b}} P(V) dV\right)
\end{align}

We can solve equation 6 numerically, assuming a thermally activated switching rate $e^{-\Delta U/kT}$, where $ \Delta U=2E_{J}(1- V_{b}/V_{c})^{\frac{3}{2}}$. A least square fit allows us to fit the histograms of Figure 3c (dashed black curves), with a dimensionless  ratio $\frac{2E_{J}}{kT}\approx 22.1$ and $\approx109$ (red and light blue histograms respectively).  Note that since $T\propto V_{rms}^{2}$, that ratio is expected to differ by a factor $\approx4.6$ between the two distributions, in close agreement with our fit. As a point of comparison, for a Josephson junction with a critical current of 1$\mu$A at 1K, this ratio would be on the order of 47. It is therefore worth noting that even though $E_{J}$, $kT$ and other energy scales in this system are many orders of magnitude larger than in a typical low-temperature Josephson junction transport experiment, the dimensionless ratios of the activation energies over $kT$ are comparable, and we therefore observe a phenomenology which is very close to what can be observed in a transport measurement.

Finally, equation 6 allows the direct calculation of $\Gamma$ using the switching histograms. Since $\Gamma\propto e^{-\Delta U/kT}$ we represent $\log(\Gamma)$ as a function of $(1-V_b/V_c)^{3/2}$ to highlight the bias dependence of the barrier height. Similar to the fits of the red histogram in Figure 3c, the linear fit to Figure 3d suggests a ratio $2E_{J}/kT\approx 22.1$, in agreement with the previously mentioned fit. 

\vspace{20mm}

\section{Inverse AC Josephson effect}

The inverse AC Josephson effect can appear when a Josephson junction is driven by an AC current \cite{Shapiro1963,Tinkham,russer_influence_2003}. The dynamics of the current driven case are not analytically solvable \cite{kautz_noise_1996}, but, since (1) is equivalent to a damped driven pendulum, qualitative descriptions rooted in this mechanical analogue offer useful insights.  In the presence of a periodic driving torque, the pendulum can enter a phase-locked state and complete $q$ revolutions during $p$ periods of the torque. This phase-locked state is resilient to small perturbations in the DC torque. The average angular frequency of the pendulum and the drive are therefore commensurate: $\langle \dot{\phi} \rangle = \frac{p}{q}\omega$. In a Josephson junction, the voltage across the junction is $\frac{\hbar\dot{\phi}}{2e}$, so phase-locking results in a quantized voltage at rational multiples of $\hbar\omega/2e$. The robustness of the phase-locked state against perturbations allows it to persist over a finite interval of the DC current, which creates voltage plateaus in the I-V curve. 

\subsection{Observation of Shapiro steps}

Although this phenomenon is typically observed when Josephson junctions are exposed to microwave radiation, in our case, the junction's frequency $f_{0}$ is a few hundreds of Hz so the AC drive can be generated by a simple lock-in amplifier.
That AC voltage is added to a DC component and supplied to the bias input $V_{b}$. The equivalent of the I-V curve which, as stated previously, is in our case a plot of $V_{out}(V_{b})$, is then expected to exhibit Shapiro steps \cite{Shapiro1963,Tinkham,russer_influence_2003}.   
\begin{figure*}[t]
    \centering
    \includegraphics[width=\textwidth]{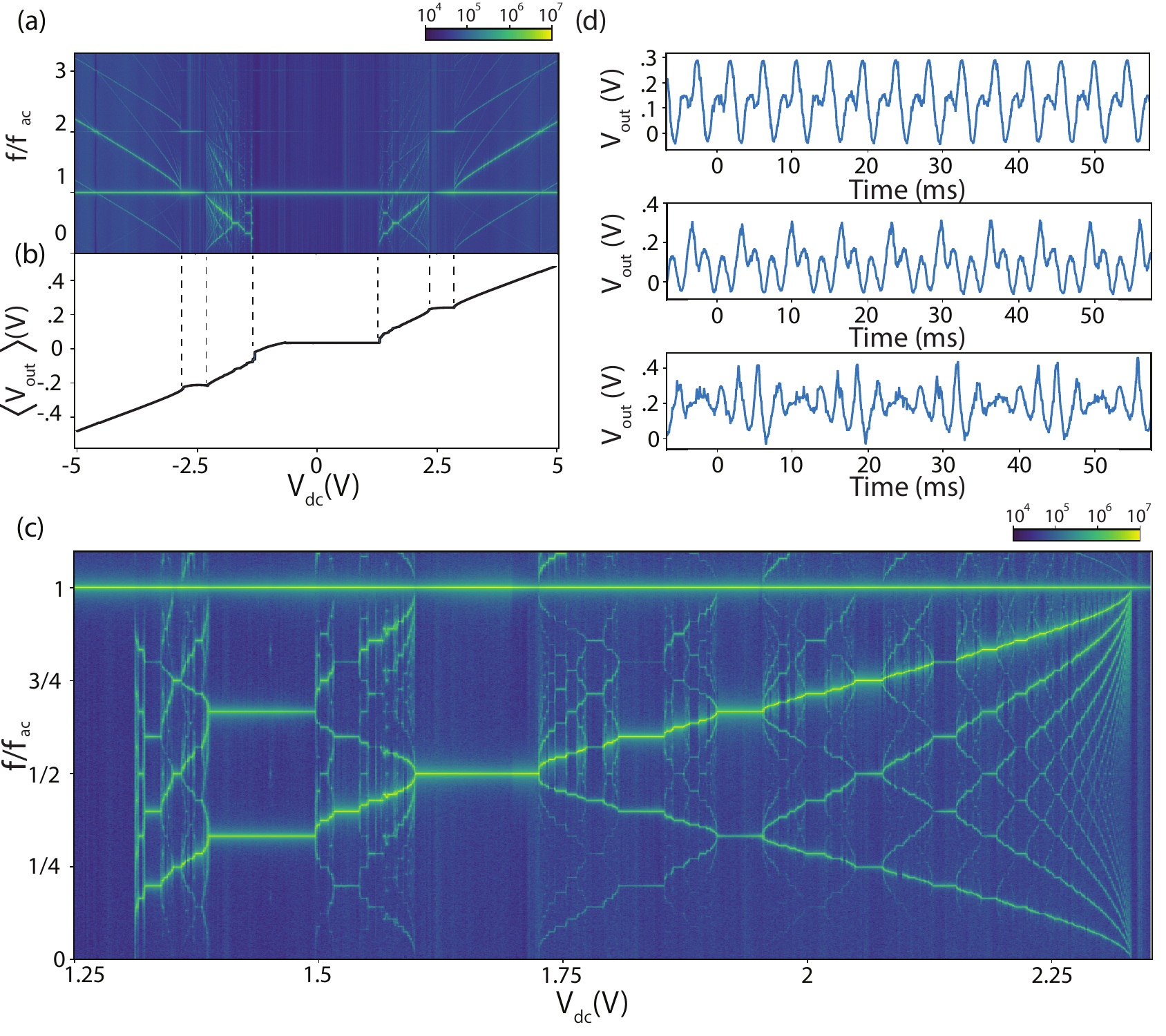}
    \caption{Fractional Shapiro steps. (a) Spectral density of $V_{out}$ as a function of $V_{dc}$ obtained by fast Fourier transform. The junction is driven by an ac voltage $V_{ac}=0.7V$. Frequencies are expressed in units of the external drive frequency $f_{ac}=457$ Hz. The junction analogue is tuned in the overdamped regime with $Q=0.4$ and $f_{0}=410$ Hz. (b) DC response $\langle V_{out}\rangle$ measured over the same range of bias values. Vertical dashed lines are guide to the eye to identify phase-locked regions. (c) Detailed map of the spectral density of $V_{out}$ as a function of $V_{dc}$ between the $n=0$ and $n=1$ integer plateaus. Measured for input biases $V_{dc}$ from 1.25V to 2.4V. (d) Examples of unfiltered time traces $V_{out}(t)$ recorded for bias values that yield subharmonic phase locking at $f_{ac}/2$ ($V_{dc}=1.65V$), $f_{ac}/3$ ($V_{dc}=1.4V$) and $f_{ac}/5$ ($V_{dc}=2.14V$)}
    \label{fig:figure5}
\end{figure*}

\par Figure 4a displays the I-V characteristics of the circuit driven by two different frequencies: one at 160 Hz and one at 280 Hz. I-V curves were collected with a fixed AC driving voltage of 0.75V and varied DC driving voltage. $\langle V_{out}\rangle$ is plotted against $V_{dc}$, and normalized in units of $f_{ac}/k$, the value of a voltage step in our setup, where k is the voltage to frequency gain of the VCO and $f_{ac}$ is the driving frequency. %This normalization factor stems from the fact that, during phase-locking, the junction voltage is $n\Phi_0 f$, where $n$ is a non-negative integer. 
This normalization helps visualize the integer quantization and shows the frequency dependence of a voltage step. Note that the different average slopes come from the normalization procedure. 
\begin{figure*}[t]
    \centering
    \includegraphics[width=\textwidth]{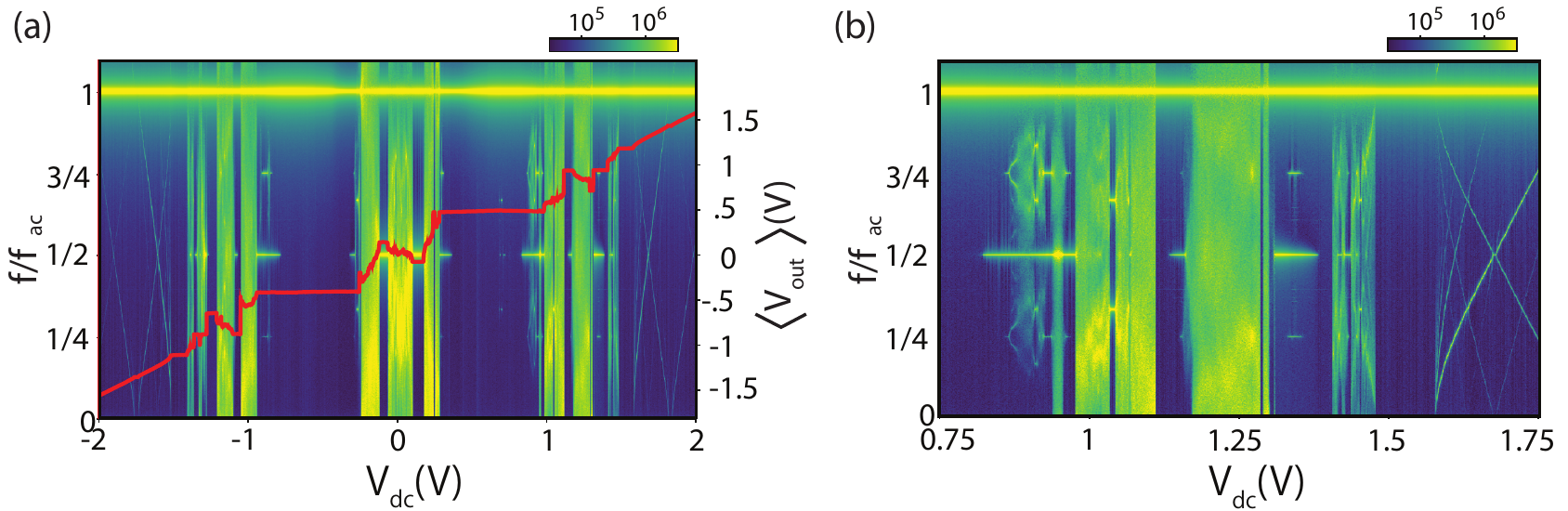}
    \caption{Chaos in underdamped analog junctions. (a) Spectral density of $V_{out}(t)$ as a function of $V_{dc}$ obtained by fast Fourier transform. An external AC drive $V_{ac}$=1.3V is applied. Frequencies are expressed in units of the external drive frequency $f_{ac}=457 Hz$. The junction analogue is tuned in the underdamped regime with $Q=1.6$ and $f_{0}=470$Hz. The DC response $\langle V_{out}\rangle(V_{dc})$ is superimposed over the map (in red) (b) Detailed map of the bifurcations between the n=1 and n=2 plateaus, using the same parameters.}
    \label{fig:figure6}
\end{figure*}
\par Figure 4b displays a map of experimentally measured differential resistance $\tilde{R}\,\equiv\,\frac{d \langle V_{out}\rangle}{dV_{dc}}$. The data was collected by measuring $\langle V_{out}\rangle (V_{ac},V_{dc})$, $V_{dc}$ being the fast axis, swept from negative to positive voltages. Both voltage axes were normalized in units of the critical voltage $V_{c}$ which was experimentally measured to be 0.73V. The resulting I-V curves were then numerically differentiated so that voltage plateaus result in dark blue regions of vanishing $\tilde{R}$, while yellow boundaries correspond to transitions between plateaus. The map bears a striking similarity to the patterns observed in conventional overdamped Josephson junction under microwave radiation \cite{larson_zero_2020}.

\par The phase dynamics of an AC driven Josephson junction can be obtained analytically when it is voltage biased, a case where the dependence of the Shapiro steps on the AC bias can be expressed with Bessel functions \cite{Tinkham, kautz_noise_1996}. In our case, the junction is current-driven and no such analytical result exists. Figure 4c was generated by numerical integration of (1), and the same averaging procedures as the experimental map of 4b were conducted. A more detailed derivation of the map 4c is given in the supplementary information. Thus, comparison of 4b and 4c allows a comparison of theoretical predictions of a differential resistance map and the results of data collected by the analog circuit. The striking similarities between the maps verifies the correspondence between theory and the circuit. 

\section{Subharmonics and chaos}

We now turn to the observation of subharmonic phase-locked states at fractional multiples of the drive frequency, when $\langle \dot{\phi}\rangle=\frac{p}{q}\omega$ and p and q are coprime \cite{benjacob_microwaveinduced_1998,kautz_noise_1996}. The more robust of those states manifest themselves as plateaus in $\langle V_{out}\rangle(V)$ at fractional multiples of the voltage quanta $f_{ac}/k$. However, we find that a direct measurement of the frequency spectrum of $V_{out}(V)$ is much more sensitive to such states. 

We set the AC excitation amplitude and frequency, and record the output frequency spectrum as a function of the bias $V_{dc}$. To that end, for each value of $V_{dc}$, we record a time trace $V_{out}(V)$ with $10^{6}$ data points spread over 2,000 to 4,000 cycles of the drive. We then compute the fast Fourier transform (FFT) for that particular time trace. Repeating this procedure at each bias value allows us to generate a map of the spectral weight as a function of $f_{ac}$ and $V_{dc}$, as shown on Figure 5a. Here, one vertical cross section of the map corresponds to the FFT of $V_{out}(t)$ at a given value of the bias. This map is measured for an analog JJ with a Q factor of 0.4, at an AC frequency of 457 Hz, and with an AC amplitude of 0.7V. Understandably, a strong peak in the frequency spectrum is seen at all biases at the fundamental drive frequency $f_{ac}$, as well as at the corresponding higher harmonics. The effective I-V curve $\langle V_{out}\rangle(V)$ is shown on Figure 5b over the same range of DC bias. We observe   n=-1, n=0 and n=+1 integer phase-locked states, which correspond to regions of the frequency spectrum where a spectral weight is only significant at the drive frequency and its higher harmonics. In the region $\abs{ V_{dc}} >2.5V$, where the I-V curve exhibits what would be a conventional Ohmic behavior in a regular junction, $V_{out}(t)$ is modulated at a frequency that increases with DC bias and is visible as a diagonal curve on Figure 5a. These correspond to quasiperiodic states which are not phase-locked \cite{kautz_noise_1996}.

The region of parameter space between phase locked states n=0 and n=1 exhibits a richer phenomenology and is shown in much greater detail on Figure 5c. A plethora of subharmonic phase-locked states are observed, as evidenced by step-like resonances in the FFT map. We show the corresponding time-domain $V_{out}(t)$ in Figure 5d for three of those subharmonics corresponding to $f_{ac}/2$, $f_{ac}/3$ and $f_{ac}/5$ (measured respectively at $V_{dc}= 1.65V$, $V_{dc}= 1.4V$, and $V_{dc}=2.14V$). 

We observed subharmonic phase-locked states for both underdamped and overdamped analog Josephson junctions. However, they are easier to see for $Q<1$ as they do not compete with chaotic states, and transitions between integer phase locked states in $\langle V_{out}\rangle(V_{dc})$ are less sharp, which leaves greater regions of parameter space where subharmonics are visible. Prior work showed that the I-V curve of a shunted Josephson junction at the onset of chaos forms a complete devil's staircase, a fractal of dimension D$\approx$0.86 \cite{shukrinov_structured_2014, yeh_fractal_1984, jensen_complete_1983,belykh_shunted-josephson-junction_1977,jensen_transition_1984,bohr_transition_1984}. Our data is reminiscent of this fractal behavior. In particular, the main resonance that goes from bottom left to the top right of the map 5c is shown in greater details in the supplementary information and shows stable plateaus at every rational number lower than 1 with denominator up to $\approx 17$. However the resolution of the data set does not allow a satisfying determination of the fractal dimension.

When the quality factor exceeds unity, $V_{out}(t)$ can exhibit chaotic behavior in addition to phase locking \cite{kautz_survey_1985,kautz_noise_1996,blackburn_experimental_1987,huberman_noise_1980,pedersen_chaos_1998,yeh_universal_1982,he_transition_1984,gwinn_intermittent_1985,cirillo_bifurcations_1982,kautz_chaotic_1981,kautz_ac_1981,jensen_complete_1983,jensen_transition_1984,bohr_transition_1984}. This is easier to see in the fast Fourier transform of $V_{out}(t)$: when chaotic, the spectral weight is significant at all frequencies so the FFT exhibits a large background as opposed to only a discrete number of peaks. Figure 6a shows an example of frequency spectrum measured as a function of the DC bias. The I-V curve $\langle V_{out}\rangle$ is superimposed in red on top of the map. Similar to Figure 5, a resonance at $f_{ac}$ dominates the spectrum of integer phase locked states visible from n=-2 to n=+2. Between plateaus, green bands corresponding to a significant spectral weight at all frequencies are visible and correspond to chaotic states. These bands are ubiquitous in those frequency spectrum maps as soon as Q exceeds 1. Transitions to chaos through period doubling bifurcations are clearly visible: they correspond to the emergence of subharmonics in the spectrum when the DC bias approaches a chaotic region (for example slightly above $V_{dc}=-1V$). Figures 6 and 5c also clearly depict intermittency. Such subharmonics appear in the spectrum before $\langle V_{out}\rangle$ starts deviating from a plateau value. This highlights the cascade of period doubling preceding chaos, as the chaotic regions of the system are located within the jumps between plateaus. Figure 6b offers a more detailed view of a cascade of bifurcations in the frequency spectrum of the same analog junction, shown only between the n=1 and n=2 plateaus. We observe an alternation of period 2, period 4 and period 3 bifurcations as well as chaotic bands. 

\begin{figure*}[t]
    \centering
    \includegraphics[width=\textwidth]{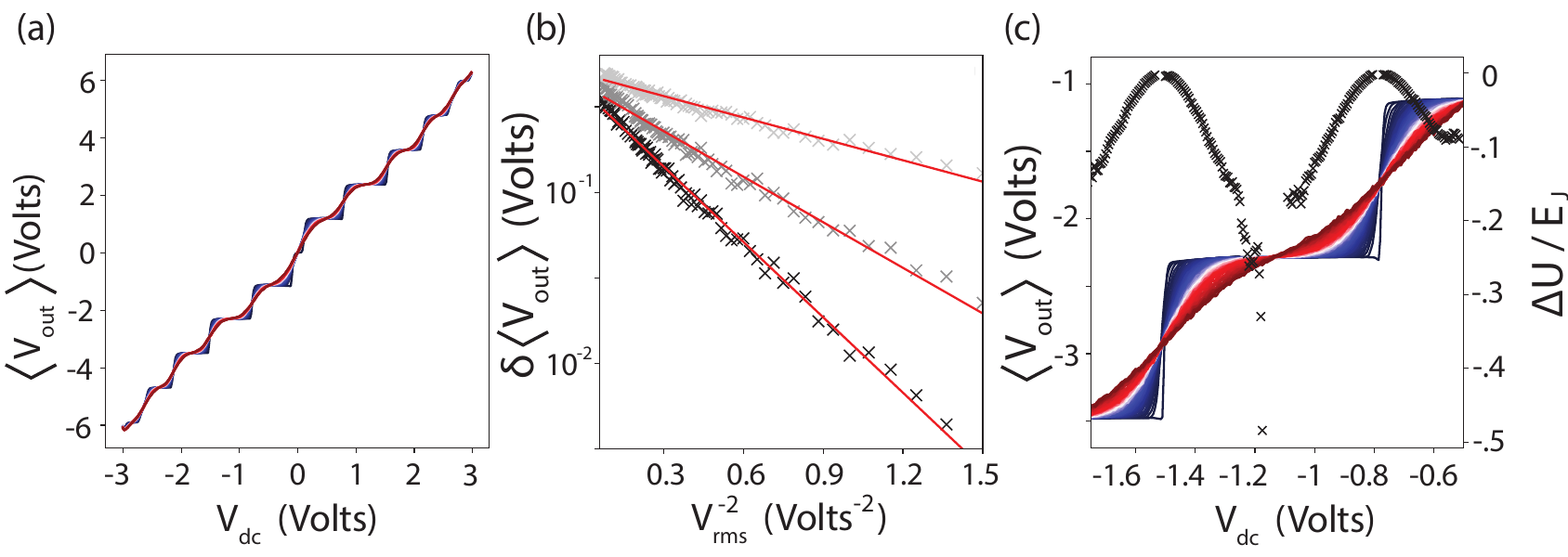}
    \caption{(a) Plot of the $\langle V_{out}\rangle$ as a function of the DC bias, in the presence of an AC excitation at 241Hz, $V_{ac}=1.35V$, and for thermal noise levels ranging from $V_{rms}=0$ to $V_{rms}=3$V. (b) Arrhenius fit of the deviation from the plateau value $\delta V$ represented in logarithmic scale as a function of $V_{rms}^{-2}$, which is proportional to $1/T$, measured at DC bias values increasingly close to the center of the plateau: $V_{dc}=-1.45V$ (light gray), $V_{dc}=-1.41V$ (dark gray) and $V_{dc}=-1.375V$ (black). (c) Activation energy as a function of bias (black).
    }
    \label{fig:figure7}
\end{figure*}
\begin{figure*}
    \centering
    \includegraphics[width=\textwidth]{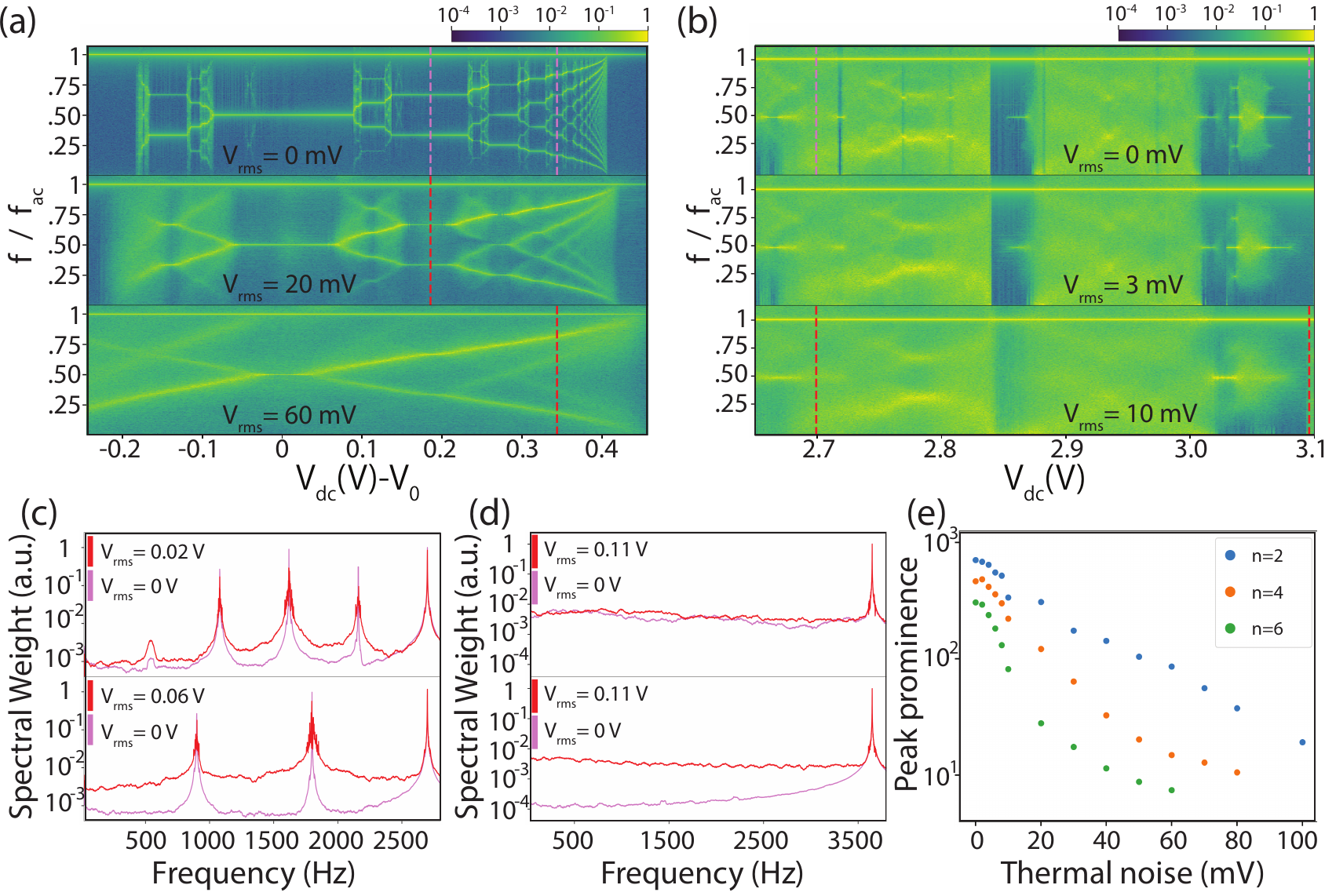}
    \caption{Temperature dependence of chaos and subharmonics. (a) Spectral density of $V_{out(t)}$ as a function of $V_{dc}$ and frequency at three different temperatures. The junction is tuned to $Q=0.93$ (R=330$\Omega$) and $f_0=430$Hz. Frequencies are expressed in units of the external drive frequency $f_{ac}=457$Hz. These maps show how thermal noise smears subharmonic resonances. (b) Spectral density maps at three different effective temperatures in the underdamped regime with $Q=1.6$, $f_0=470$, $f_{ac}=457$Hz, and DC bias voltage swept from 2.65 to 3.1V. Notice how the background noise increases with temperature, but leaves chaotic regions unchanged. (c) Spectral weight versus frequency line cuts of the subharmonic plateaus from panel (b). These more explicitly show the decreasing peak height and increasing background seen in panel (b). (d) Line cuts of chaotic maps from panel (a) showing spectral weight versus frequency. Notice that as opposed to the subharmonic structure in panels (b) and (c), thermal noise does not affect chaos. (e) Peak prominence of subharmonic plateaus versus $V_{rms}$ temperature. The prominences are measured as the ratio between the average peak height of the subharmonics and the background average.
}
    \label{fig:figure8}
\end{figure*}

\subsection{Thermal noise in the AC driven analog junction}
\par

Our experimental setup allows us to probe the stability of phase-locked and chaotic states in the presence of thermal noise. 

We first discuss integer phase-locked states. The circuit is first tuned to a quality factor of approximately 0.6 and a junction frequency of 450 Hz. An AC bias is applied to the circuit at a frequency of 241Hz, a large fraction of $f_{0}$. In what follows the bias amplitude is fixed at a value of 1.35V. The time-averaged output voltage of the circuit $\langle V_{out}\rangle$ then shows Shapiro steps as a function of the DC bias, as shown on Figure 7a. 

In addition to the AC and DC biases, we apply a random voltage source of variable amplitude to the circuit. We again use the fact that the effective temperature corresponding to the noise source scales like $V_{rms}^{2}$. The resulting I-V curves are shown on Figure 7a at different effective temperatures, ranging from zero to $V_{rms}=3V$. Transitions between plateaus become more rounded as the effective temperature increases. 

The deviation from the plateau value $\delta V$ is expected to be thermally activated as $\delta V\propto e^{-\Delta U/kT}$, where $\Delta U$ is the quasipotential quantifying the stability of the phase locked state. We thus plot the error voltage $\delta V$ in logarithmic scale as a function of $V_{rms}^{-2}$, which is proportional to the inverse effective temperature (Figure 7b). The thermal activation behavior is evident from the linearity of the Arrhenius plot over more than one order of magnitude. That trend is represented for three different DC bias values ($V_{dc}$=-1.45, -1.41 ad -1.375V) at bias values that get closer to the center of the n=2 plateau. The activation energy gets larger closer to the center of plateau, indicating an increased stability of the phase locked state. 
We turn to a more systematic representation of the quasipotential $\Delta U$ as a function of the bias. Figure 7c represents the I-V curves $\langle V_{out}\rangle (V_{dc})$ at increasingly large effective temperatures (blue to red curves). This data set allows us to determine the activation energy scale $\Delta U$ as a function of the DC bias for a given phase locked state. It is represented in black and in units of the Josephson energy for the circuit. One can see how $\Delta U$, which quantifies the stability of the phase locked state, vanishes at the boundaries of the plateau. Meanwhile $\Delta U$ is largest at the center of the plateau, where the phase locked state is most robust. Very close to the center of the plateau ($V_{dc}\approx -1.1$V), the high temperature traces intersect the middle of the low-temperature plateau. The deviation $\delta V$ therefore becomes vanishingly small even at high temperature, and it is therefore experimentally unfeasible to extract the energy scale $\Delta U$ for the few points closest to that plateau center.
Finally, note that the quasipotential is expressed in units of the Josephson energy; the procedure we followed to make this conversion is detailed in the supplementary information. 

In the case of fractional phase locked states the error voltage $\delta V$ between low and high temperature is much smaller, which prevents us from obtaining a satisfying Arrhenius fit to $\delta V (T)$. We thus turn to a determination of the fast fourier transform of $V_{out}(t)$ at different effective temperatures. Figure 8a represents this FFT as a function of the DC bias voltage $V_{dc}$ and the frequency in units of $f_{ac}$. It is measured between the n=1 and n=2 Shapiro steps, in a region of parameter space where numerous subharmonic steps are visible (the circuit is tuned so that $Q=0.93$, $f_{0}=430$Hz and $f_{ac}=457$Hz). Similar to previous FFT maps, a resonance at the excitation frequency is visible in all maps, and additional resonances at rational multiples of $f_{ac}$ indicate fractional phase locking. Three panels are measured at increasing effective temperatures, which are controlled by the RMS amplitude of the voltage noise. We see how fractional resonances are smeared by thermal noise, with low denominator fractions being the most robust (the period-doubling resonance at $f_{ac}/2$ is the last to survive). 

To better visualize the smearing of fractional phase locked states, we show on Figure 8c cross sections of the FFT as a function of frequency at two different effective temperatures. Cross sections are shown for subharmonics $f_{ac}/3$ and $f_{ac}/5$. The FFT are normalized so that the peak amplitude at the drive frequency has a weight of 1. Evidently, the background of the FFT rises with temperature, which causes the prominence of the FFT peak to drop with temperature. This prominence is shown as a function of thermal noise in Figure 8e. We defined the prominence as the average of the peak amplitudes of the FFT at multiples of $f_{ac}/n$, divided by the average of the background FFT value. This prominence is seen to drop with thermal noise, with a steeper drop for the larger denominator subharmonic (n=5). This reflects the disappearance of the phase-locked state as the $V_{out}(t)$ becomes noisier.

We turn to the dependence of the chaotic bands on the thermal noise level. Similar to Figure 8a, in Figure 8b we represent the FFT of $V_{out}(t)$ as a function of $V_{dc}$ at different effective temperatures. Here, the circuit is tuned to be underdamped with $Q=1.6$, $f_0=470$Hz, $f_{ac}=457$Hz and $V_{ac}=0.9V$ which allows for chaotic behavior to arise at certain bias values. As opposed to the fractional resonances, these chaotic bands do not depend on temperature within the range of applied noise levels. This distinction is most clear by comparing how the phase-locked regions smear while the chaotic regions remain essentially unchanged. To illustrate this, we take cross sections of these maps in Figure 8d just as we did for Figure 8c. The top map is taken in a chaotic region and we can see that the background does not noticeably change at higher effective temperatures ($V_{rms}$=0.11V). The bottom map was taken when only the fundamental frequency was present and as temperature is increased the background increases which causes the prominence of the peak to drop.

Our results illustrate how this analog circuit replicates a wide range of Josephson junction phenomena. In the underdamped case, the retrapping voltage scales like the inverse of the quality factor, and the effective temperature dependence of the  switching voltage distribution closely matches expectations from the RCSJ model. In overdamped juntions, we observe the diffusion of the phase in the presence of thermal noise.
The addition of an AC excitation allows us to observe chaotic states, integer and fractional phase locked states, and their dependence on AC power and temperature. The circuit provides a simple setup to predict phase dynamics and help design circuits based on Josephson junctions. In particular, it paves the way toward the implementation of more complex analog circuits emulating the phase dynamics of driven multiterminal Josephson junctions \cite{Draelos2019, Pankratova2020, Pribiag2020}, which recently attracted a lot of interest due to the novel quantum and topological phenomena they can help engineer \cite{Riwar2016}. 

\section{Acknowledgments}

We thank Stephen Teitsworth, Ethan Arnault, Trevyn Larsson, Lingfei Zhao and Gleb Finkelstein for helpful discussions. Sara Idris and Aeron McConnell acknowledge funding from the Graduate Research Assistantship Mentoring Program and the Student And Faculty Excellence Fund at Appalachian State University.
\bibliography{references}
\clearpage

\onecolumngrid

\begin{center}
    {\LARGE \textbf{Supplementary information}}
\end{center}

\begin{figure}[b]
    \centering
    \includegraphics[width=\textwidth]{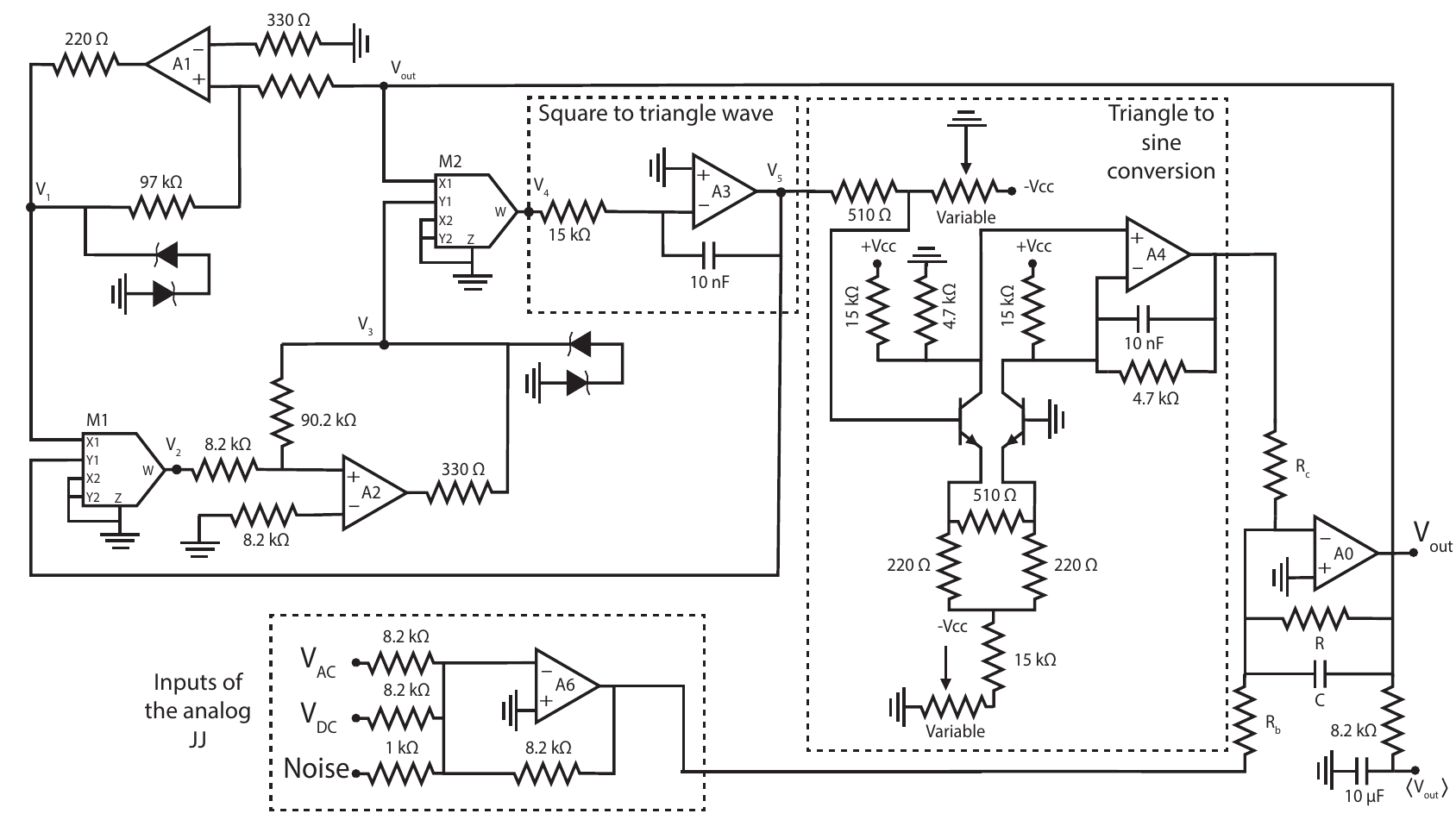}
    \caption{Schematic of the analog Josephson junction circuit used in the main paper. A0 corresponds to the operational amplifier shown in the main text.
    }
    \label{fig:figure9}
\end{figure}
\twocolumngrid
\vspace{5mm}
\section{Detailed circuit for the voltage controlled oscillator}

Figure 9 is a diagram of the full circuit used in this paper. Some of the main sections, including the triangle to sine converter and input voltage adder, are indicated for clarity. The part of the circuit that generates the triangle wave was proposed by Ref. \cite{blackburn_circuit_2007}

%\end{widetext}

The junction's output voltage $V_{out}$ corresponds to the input of the voltage controlled oscillator. Two back to back Zener diodes with a breakdown of approximately 10V are placed at the output of A1. Given that the inverting terminal of A1 is held at ground potential, $V_{1}$ is therefore sign($V_{out}$)$\times$10V. The multiplier M1 has a gain of $\frac{1}{10}$, so its output $V_{2}$ is equal to sign($V_{out}$)$\times$$V_{5}$, where $V_{5}$ is the triangular wave output voltage, which is fed back to one of the inputs of multiplier M1. 

That signal is then fed to a Schmitt trigger (A2), so that voltage $V_{3}$ is equal to $\pm$ the Zener breakdown voltage (10V) and flips sign whenever the triangle wave output reaches its maximum amplitude, which is $\frac{8.2\times V_{Z}}{90.2}$. The voltage at the output of multiplier M2 is equal to $V_{out}$ so the rate of change of the triangle wave output is $\pm\frac{V_{out}}{\tau}$, where $\tau$ is the time constant of the integrator circuit. The frequency of the triangle wave is thus proportional to the input of the VCO ($V_{out}$).   

The triangle wave voltage $V_{5}$ is fed to a differential amplifier pair of BJTs whose output approximate a sine wave by a piece-wise hyperbolic tangent signal. We show in Figure 11 how this signal has heavily suppressed higher harmonics. 

\section{Derivation of the main differential equation}

We now discuss how the dynamical properties of the circuit shown on Figure 1a are described by equation 1. Assuming a vanishing input current on the inverting terminal of the operational amplifier shown on Figure 1a, one can write:
\begin{equation}
    \frac{V_{b}}{R_b}+\frac{\alpha}{R_c}\sin(2\pi k\int V_{out}\,\,dt)+\frac{V_{out}}{R}+C\frac{dV_{out}}{dt}=0
\end{equation}

We define $\phi\equiv2{\pi}k \int V_{out} \,dt$ which yields: 

\begin{align}
    \frac{-V_{b}}{R_b} = \frac{{\alpha}}{R_c}\sin{\phi} + \frac{\dot {\phi}}{2{\pi}kR}+ \frac{C}{2{\pi}k}\ddot {\phi}
\end{align}.

We define the junction's frequency as:   
${\omega}_o^2 \equiv\mathlarger{ \frac{2{\pi}{\alpha}k}{R_cC}}$, the quality factor $Q=RC\omega_{0}$, and the critical voltage $V_{c}=-\frac{\alpha R_b}{R_c}$. This allows us to rewrite equation (8) as:

\begin{align}
\ddot {\phi} + \frac{\omega_{0}}{Q} \dot {\phi} + \omega_{0}^{2}\sin{\phi}=\omega_{0}^{2}\frac{V_{b}}{V_{c}}
\end{align}

Assuming $V_{b}$ is constant, a first integral of equation (9) can be written as:

\begin{align}
\frac{d}{dt}\left(\frac{\dot{\phi}^{2}}{2}-\omega_{0}^{2}\cos(\phi)-\omega_{0}^{2}\frac{V_{b}}{V_c}\phi\right)=-\frac{\omega_{0}\dot{\phi}^{2}}{Q}
\end{align}

Using $\dot{\phi}=2\pi k V_{out}$, this becomes:

\begin{equation}
\frac{d}{dt}\left(\frac{1}{2}C V_{out}^{2}+U(\phi)\right)=-\frac{V_{out}^{2}}{R}
\end{equation}

We defined the washboard potential $U(\phi)$ as:

\begin{align}
U(\phi)=-E_{J}\left(\cos(\phi)+\frac{V_{b}}{V_c}\phi\right)
\end{align}

Using $E_J= \frac{\alpha}{2{\pi}k\,R_c}$. The right hand side of equation (11) corresponds to the dissipation through resistor $R$. The capacitor's energy plays the role of a kinetic term.

\subsection{Numerical integration of the differential equation}

We start from the following equation:

\begin{equation*}
\ddot {\phi} + \frac{\omega_{0}}{Q} \dot {\phi} + \omega_{0}^{2}\sin{\phi}=\omega_{0}^{2}\frac{V_{b}}{V_{c}}
\end{equation*}

We rewrite this second-order ODE into a coupled system of first-order ODE's.

\large
\begin{equation*}
\begin{cases}
\xi\,=\,\dot{\phi}\\
\dot{\xi}\,+\,\frac{\omega_0}{Q}\xi\,=\,\omega_0^2\frac{\left(V_{dc}\,+\,V_{ac}\cos(\omega t)\right)}{V_c}-\omega_0^2\sin(\phi)
\end{cases}
\end{equation*}
\normalsize

Note here that $V_b\,=\,V_{dc}\,+\,V_{ac}\cos(\omega t)$. 

%Once the parameters are set, two empty vectors for $V_{ac}$ and $V_{dc}$, with appropriate lengths for the map dimensions, and an $m\times n$ empty array- from here denoted $V_{out}$- were created based on the dimensions of $V_{dc}$ by $V_{ac}$. Following this, a double-nested for loop was run, where the primary index was the $n$th $V_{ac}$ value and the nested for loop ran across all $m$ values of $V_{dc}$. Thus, for each $n$th iteration, $V_{ac}$ remained fixed, while each $V_{dc}$ value is iterated through. For each $n$th iteration, a temporary $m\times p$ array was generated, where $p\,=\,10^5$. 
We solve this system of equation with a 4th order Runge-Kutta method which is parallelized with the python package Numba. Each numerical integration of $\phi(t)$ is run over a total time of $500T$, with a time step of $\frac{T}{200}$. The array of $\dot\phi$ values was obtained and multiplied by a factor of $\frac{1}{2\pi K_J}$ to convert to $V_{out}$. When necessary, $V_{out}(t)$ is time-averaged over the last 400 periods to determine the DC component (i.e. the plateau value in the case of Shapiro steps). 
\begin{figure}
    \centering
    \includegraphics[width=0.5\textwidth]{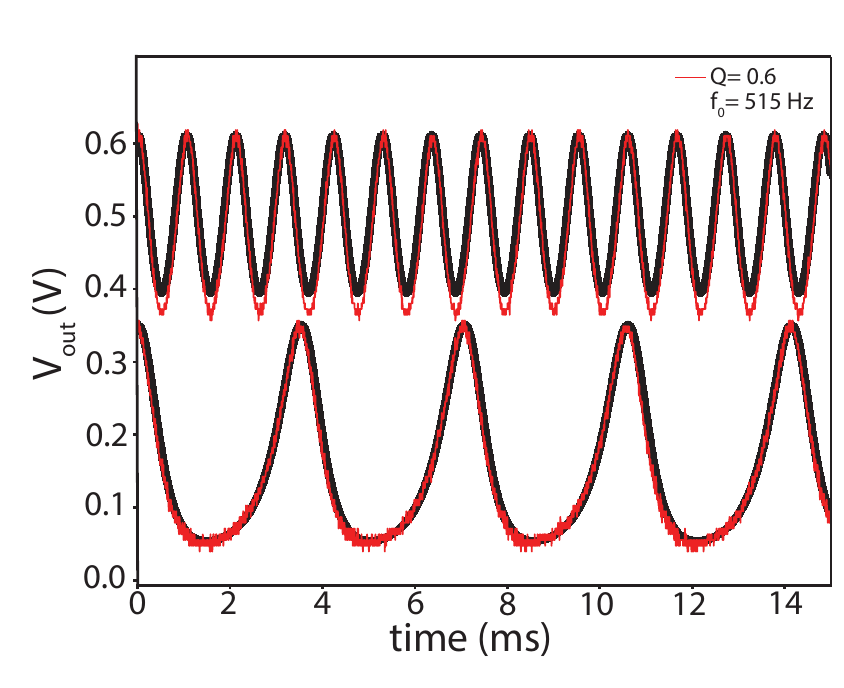}
    \caption{Unfiltered output $V_{out}(t)$ of the DC driven analog junction in the resistive state.  This is proportional to the time derivative of the phase as it descends the washboard potential (see Figure 1b). When greater bias voltage is applied, not only is the DC component of the output increased (top curve) indicating an overall higher average phase velocity, but the period is also impacted. As the bias voltage is increased the period decreases indicating that the phase point is traversing full cycles of the washboard more quickly. While the bias is decreased the period diverges as it approaches the transition to the zero-voltage state.}
    \label{fig:figure10}
\end{figure}

As an example of the fit between our numerical simulation and our experimental observations, we show on Figure 10 the output voltage $V_{out}(t)$ before time-filtering (in black) compared to the Runge-Kutta derivation (in red), showing excellent agreement between the two. 

\begin{figure*}
    \centering
    \includegraphics[width=\textwidth]{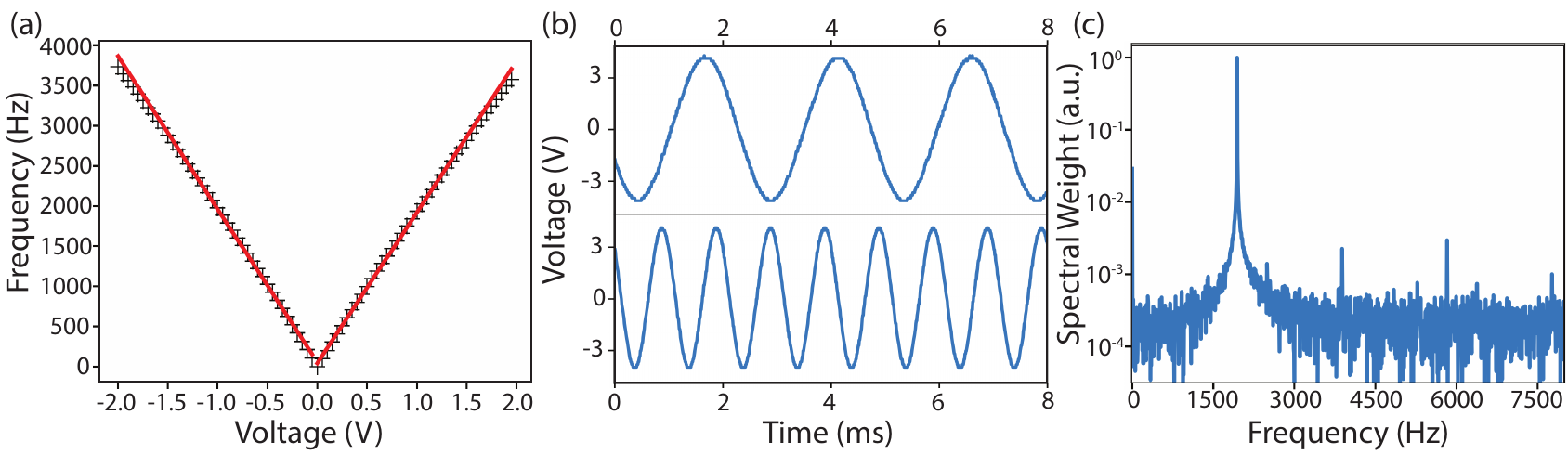}
    \caption{Characterization of the VCO. (a) 80 voltages ranging from -2V to 2V were supplied and the output frequency was measured and plotted against these inputs. The data was fit using linear models showing that our VCO does indeed create a linear relationship between input voltages and output frequencies. (b) The output sine waves are shown for 0.2V (top) and 0.5V (bottom). (c) Taking the Fourier transform of one of the output sine waves we can see that the higher harmonics are suppressed by multiple decades.}
    \label{fig:vco_figure}
\end{figure*}
\section{Characterization of the VCO}

Figure 11 highlights some of the main properties of the voltage controlled oscillators. The linear relationship between the input voltage of the VCO and its output frequency is shown on Figure 11a. A fit, shown in red, allows us to extract a voltage to frequency gain of approximately 1900 Hz.V$^{-1}$. Two examples of the VCO output voltage are shown in Figure 11b for input voltages of 0.2 and 0.5V respectively. Our circuit includes a triangle to sine converter whose purpose is to suppress the higher harmonics initially present in the triangle wave output of the integrator. Figure 11c shows the fast Fourier transform of the output of the VCO (in log scale) and confirms that the higher harmonics are heavily suppressed since there are more than two decades of suppression between the primary frequency and the higher harmonics.
\begin{figure}
    \centering
    \includegraphics[width=0.5\textwidth]{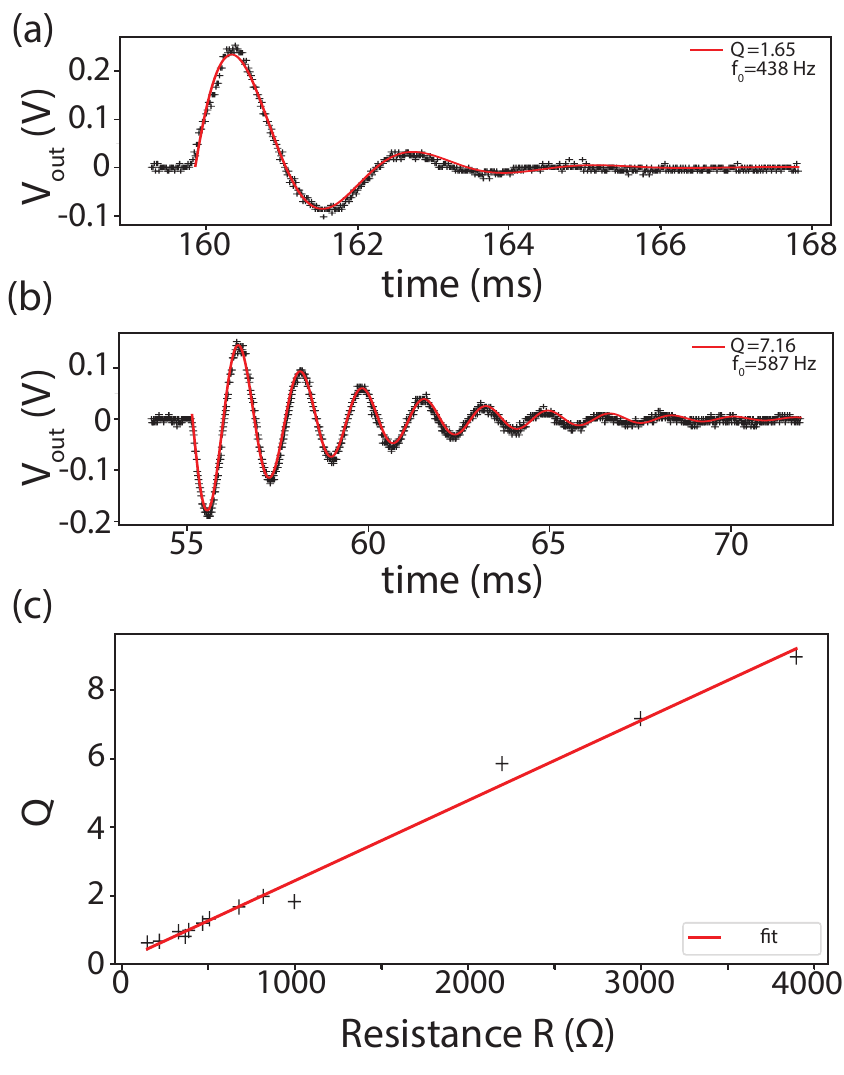}
    \caption{a) Oscillation in $V_{out}(t)$ following a step function in the input $V_{b}$. $V_{out}(t)$ decays down to zero, similar to the angular velocity of a pendulum following an impulse. The Q factor of the circuit is extracted from the fit and equal to 1.65. b) Similar measurement for Q=7.16. c) Quality factors plotted as a function of the shunting resistance R, showing a roughly linear dependence.  }
    \label{fig:vco_figure}
\end{figure}
\section{Determining the Q factor of the analog junction}

Recall the differential equation (1) that characterizes the circuit. When the applied bias voltage is a step function, the behavior of $V_{out}(t)$ following a falling edge of $V_b$ can be approximated as that of an exponentially decaying sine. Within the small $\phi$ approximation, the output voltage versus time curve can be fit to the equation below, and thus, the quality factor Q and frequency $f_0$ of the circuit can be determined.

\begin{equation*}
\begin{cases}
V_{out}(t)=A.e^{-\beta t}\sin(\omega (t-t_{0}))
\\
\omega^{2}\,\equiv\,\omega_0^{2}\,-\,\beta^{2},    \,\beta\,\equiv\,\frac{\omega_0}{2\,Q}\\

\end{cases}
\end{equation*}
Figure 11a shows the data for a resistor R of 680\,$\Omega$. When fit successfully to the above equation, as indicated by the red curve, Q is found to be 1.65 and $f_0$ is 438\,Hz. Similarly, when R=\,3\,k$\Omega$, the fit shown in figure 10b reveals a Q of 7.16 and an $f_0$ of 587\,Hz. To confirm that the quality factor is linearly related to the resistor value as the expression $Q\equiv RC\omega_{0}$ suggests, the procedure shown in figures 12a and 12b was repeated for multiple values of the resistor R. Plotting R vs Q as well as the linear fit for R values less than 4\,k$\Omega$ resulted in figure 12c. We see the expected relationship between R and Q values. At very large values of R, corresponding to Q$>$10, this linear relationship is no longer accurate.

\section{Determination of the quasipotential in units of $E_{J}$}

In Figure 4f, the energy scale for the thermal activation out of a phase locked state is expressed in units of the Josephson energy $E_{J}$. 

To do this we first perform an Arrhenius fit to data comparable to Figure 4e, but acquired for a range of DC bias values. This yields the following functional form for the error voltage as a function of the RMS voltage noise: 
\begin{equation}
\delta\langle V_{out}\rangle\propto exp\left(-\frac{\alpha}{V_{rms}^{2}}\right)=exp\left(-\frac{\Delta U}{k_{B}T_{eff}}\right)
\end{equation}

We use the Johnson-Nyquist noise expression to relate the voltage noise amplitude to the effective temperature, and find that the quasipotential $\Delta U$ is related to the Arrhenius fitting coefficient $\alpha$ as: $\Delta U=\frac{\alpha}{4BR}$. $B$ represents the bandwidth of the junction and $R$ its shunting resistance.  

We can eliminate those two parameters and express them as a function of the Josephson energy $E_{J}$ using a measurement of the zero bias differential resistance caused by phase diffusion in the absence of AC excitation: 

\begin{equation}
\frac{d\langle V_{out}\rangle}{dV_{dc}} \propto \frac{1}{T}exp\left(-\frac{\alpha '}{V_{rms}^{2}}\right)=\frac{1}{T}exp\left(-\frac{2E_{J}}{k_{B}T_{eff}}\right)
\end{equation}

$\alpha '$ is experimentally determined and relates to the Josephson energy as: $\frac{\alpha '}{8BR}=E_{J}$. 

Using this relation we can therefore eliminate $B.R$  from the quasipotential expression and express $\Delta U$ in units of $E_{J}$.

\section{Different representation of subharmonics}

We showed in Figure 5 of the main paper that the analog junction exhibits fractional phase locking, which results in resonances in the fast Fourier transform maps at rational multiples of $f_{ac}$. An alternative representation of these resonances is to determine for each DC bias value the peak frequency in the FFT (strictly lower than the drive $f_{ac}$). This yields the curve shown in Figure 13, which is obtained from the same data as Figure 5. Panel b is a magnified plot corresponding to the red dashed box in panel a. On both plots, we show a few of the rational reduced frequencies within the measured interval.   

\begin{figure}
    \centering
    \includegraphics{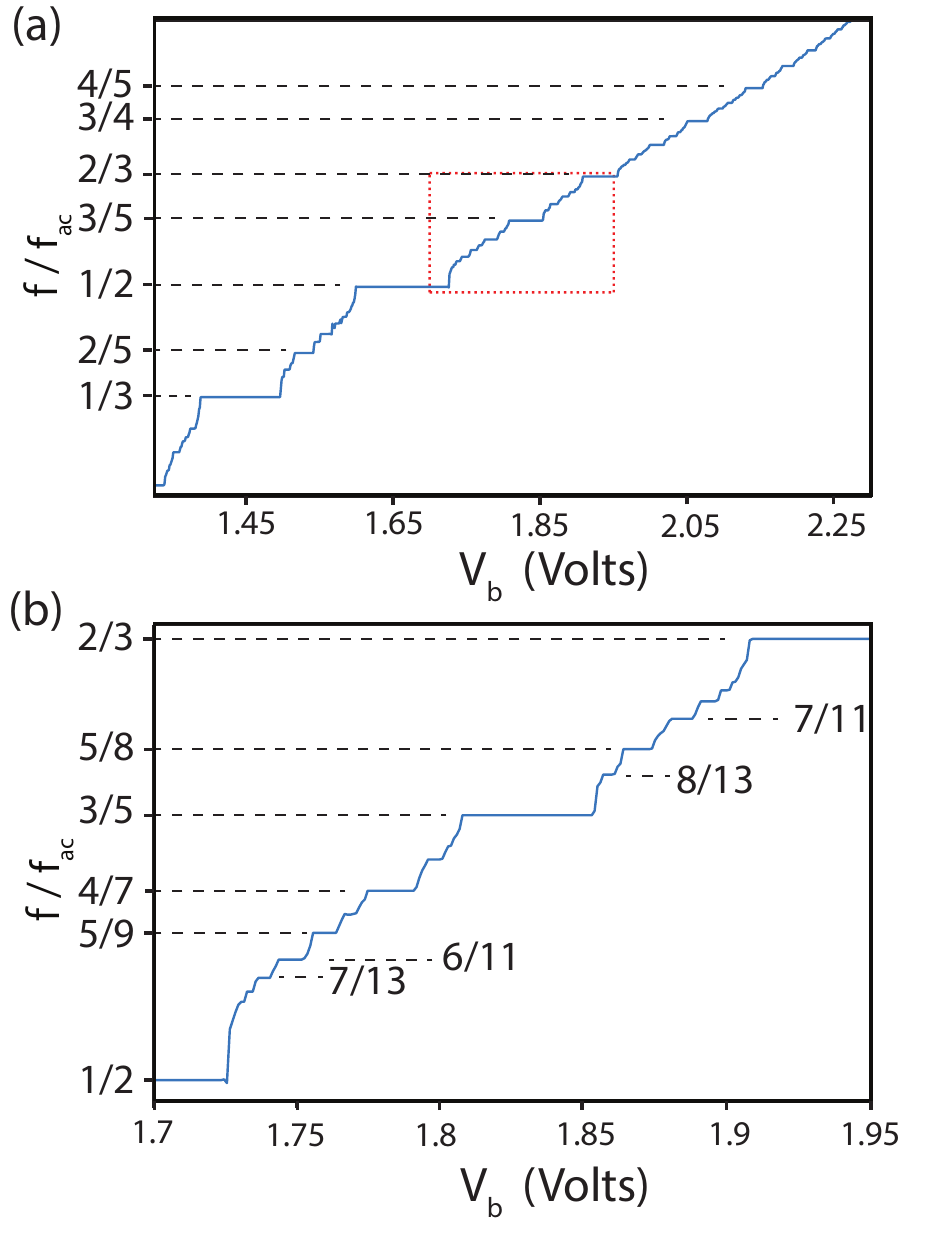}
    \caption{a) Dominant subharmonic frequency in the spectrum of Figure 5c, shown as a function of DC bias $V_{b}$. b) Magnified plot of the same data, corresponding to the red dashed box in panel (a).}
    \label{fig:my_label}
\end{figure}
\end{document}